\documentstyle[aps,epsfig]{revtex}
\begin{document}
\title{Centrality  dependence  of high $p_T$ suppression in Au+Au
collisions: dynamical or geometric effect?}

\author{\bf A. K. Chaudhuri\cite{byline}}
\address{ Variable Energy Cyclotron Centre\\
1/AF,Bidhan Nagar, Kolkata - 700 064\\}
\maketitle
\begin{abstract}
In a pQCD-based model, we have analyzed the STAR and PHENIX data on the high 
$p_T$  suppression  of  charged  hadrons and neutral pions,  in  Au+Au collisions at $\sqrt{s}$=200 GeV. In the  jet  quenching  or  the  energy  loss picture,  $p_T$ dependence of the nuclear modification factor of the charged hadrons and neutral pions, in  all  the  centrality  ranges,  are   well explained in terms of a single parameter, the average relative energy loss $<\Delta E/E>$. $<\Delta E/E>$  increases with the centrality, from a small or near zero value in the most peripheral collisions, it increases to $\sim$ 20-22\% in most central collisions. However, a part of the increased energy loss with centrality, is due to  geometric effect, i.e.  increased path length in central than in peripheral collisions. If the geometric part of the energy loss is excluded, the extracted energy loss reveals that even in central collisions, the density of matter is not very high. It is only 1.5-2.0 times the normal nuclear matter density. Our analysis thus precludes formation of high-density matter in central Au+Au collisions.
\end{abstract}   

\pacs{PACS   numbers:    25.75.-q, 13.85.Hd, 13.87.-a}

\section{Introduction}

One  of the predictions of quantum chromodynamics is the possible
existence  of  a  deconfined  state  of  quarks  and  gluons.  In
relativistic  heavy  ion  collisions,  under  certain  conditions
(sufficiently  high  energy  density  and  temperature)  ordinary
hadronic  matter  (where  quarks  and  gluons  are  confined) can
undergo a phase transition to a deconfined matter, commonly known
as quark gluon plasma (QGP). Over the years,  nuclear  physicists
are  trying to produce and detect this new phase of matter, first
at CERN SPS and now at Relativistic Heavy  Ion  Collider  (RHIC).
At RHIC several new results are obtained, most important among
them is the high $p_T$ suppression in Au+Au collisions and its absence in d+Au collisions. All the four
experiments, STAR, PHENIX, PHOBOS and BRAHMS \cite{star1,phenix1,phobos1,brahms1} observed that high $p_T$
hadron production  in  Au+Au  collisions at 
$\sqrt{s}=200$ GeV (as well as at $\sqrt{s}=130$GeV), {\em  scaled}  by  the
average  number  of binary collisions, is suppressed with respect
to pp collisions. The  suppression is more  in  central  collisions
than  in  peripheral  collisions.  
The observation has created an excitement in nuclear  physics
community, as a possible signature, if not of QGP, of creation of
highly dense matter in central Au+Au collisions.

One  of the explanations offered for the high $p_T$ suppression at
RHIC is the jet quenching \cite{gy90,wa92}. 
In  the  fragmentation
picture   \cite{co82},   the   single   parton  spectrum  is
convoluted with the probability for a  parton  $i$  to  hadronize
into  a  hadron  $h$,  which  carries  a  fraction $z < 1$ of the
momentum of the parent parton. Unlike in  pp  collisions,  in  AA
collisions,  partons  produced  in the initial hard scattering 
has to propagate through a medium before fragmentation. While
propagating through the medium,
the partons can suffer 
energy loss  through multiple scattering.
As  a  results,  at  the  time  of
fragmentation to  hadrons,  the  partons will  have  less  energy  and
consequently will produce  less  hadrons  than would have otherwise.
Due to increased path length, partons will suffer more 
energy loss in central collisions than in peripheral collisions.
The jet quenching or energy loss picture thus predicts more
high $p_T$ suppression is central than in peripheral collisions,
in agreement with the experimental observation.

Partonic energy loss in a medium  is a complex
phenomena. Partons lose energy mainly due to gluon radiation.
Recently, much progress has been made in the calculation of 
radiative energy loss of partons in medium, by several 
authors and in different approaches \cite{gy94,ba97,ba00,gy00,gy00a,we00,gu00,wa02}. However, exact nature
of the energy loss is still uncertain.
For example, in the BDMS \cite{ba97,ba00} relative energy loss increases rapidly with decreasing energy, $\Delta E/E \propto 1/E$, but in  GLV \cite{gy00}
relative energy loss is approximately constant. However, all the models agree that in a 
{\em static} medium, energy loss due to gluon radiation 
have a characteristic quadratic dependence on the in medium
path length. In an one dimensionally expanding medium, the quadratic dependence is diluted to a linear dependence \cite{gy00a}.

In the jet quenching or the energy loss picture, several authors 
have studied high $p_T$ production at RHIC energy in Au+Au
collisions \cite{wa02,wa96,je03,mu03,vi02,wa03,vi04}. The LO or
NLO pQCD models with jet quenching can well explain the high
$p_T$ suppression, implying large parton energy loss or high initial gluon density. 
In \cite{wa02} gluon density,
at an initial time  $\tau_0$= 0.2 fm was estimated to be 15 times higher than that 
in a normal nuclear matter.

Apart from the jet quenching, alternative explanations are also offered
for the high $p_T$ suppression. Indeed, leading hadrons from the
jet fragmentation can possibly have strong interactions with the
medium created and possibly be absorbed or its $p_T$ be shifted to lower values leading to the high $p_T$ suppression. In \cite{ga03} high $p_T$ suppression,  due to final state hadronic
interactions, was considered. It was argued that partons cannot materialize into hadrons in a deconfined phase. Fragmentation
can occur only outside the deconfined phase or in vacuum. It was argued that
high $p_T$ (pre-)hadrons 
can well be realized inside the late stage of the fireball. Then interaction of the (pre-)hadrons with bulk hadronic matter could
lead to the observed suppression. RHIC data on high $p_T$ suppression can be partially explained in the model.
Final state hadronic interaction was also used by 
Capella et al\cite{ca04} to explain the high $p_T$ suppression. In their model, high
$p_T$ hadrons interact with the comovers and their $p_T$ is shifted to smaller
values.  
Due to steepness  of the $p_T$ distribution the effect could be large. Indeed, it was shown that RHIC data are well explained
in such a model \cite{ca04}.  

Recently Xin-Nian \cite{wa03} has argued strongly against models 
using the final state hadronic interaction to explain RHIC high $p_T$
suppression.
He argued that hadron formation
time could not be small, rather large, in the range of 30-40 fm,
much larger than the typical medium size or lifetime of the dense medium. Moreover, nearly flat $p_T$ dependence of the observed
suppression at high $p_T$ empirically leads to a linear energy
dependence of the hadronic energy loss. Since the hadron
formation time is proportional to hadron or jet energy, the total
 energy
loss due to hadron rescattering or absorption decreases with energy. He argued that observed high $p_T$ suppression at RHIC
could only be due to partonic energy loss.

As told earlier, jet quenching or partonic energy loss does explain the observed
high $P_T$ suppression. In the jet quenching picture, one tries to deduce the
density of the medium produced and also tries to comment on the question of
thermalization of the medium. 
However, to comment about the medium
where the partons undergo energy loss, careful analysis is required to eliminate effects due to geometry of collisions.
In the present
paper we have analyzed three data sets on high $p_T$ suppression, (i) STAR data on charged particle production, (ii)
PHENIX data on neutral pions and (iii) PHENIX data on charged hadrons.
Considering the existing differences in model calculations of energy loss,
we took a simplistic approach. We assume that in a
certain centrality range, partons suffer, on the average, 
$<\Delta E>$ energy loss. Then treating the average energy loss as a parameter, we fit the experimental data. As will be shown below, both the STAR and PHENIX data
on nuclear modification factor, in all the centrality ranges, are well explained in
terms of a single parameter, the average relative energy loss $<\Delta E/E>$.
From the centrality dependence of the extracted relative energy loss we then try to learn about the medium in which the partons suffered the energy loss. 

The plan of the paper is as follows:
In section 2, we have described the pQCD model for hadron production. In section 3, results of our analysis of STAR and PHENIX data
are presented. Summary and conclusions are given in section 4.

\section{pQCD model for hadron production}

Details of high $p_T$ hadron production in pp collisions could be
found  in  \cite{eskola03}. In pQCD model, the differential cross
section for production of a hadron  h  with  transverse  momentum
$q_T$  ,  at  rapidity  $y$, in pp collisions can be written as ,

\begin{eqnarray} \label{1}
\frac{d\sigma^{pp->hX}}{dq^2_Tdy} = &&K J_(m_T,y) \int \frac{dz}{z^2}
\int dy_2  \sum_{<ij>,<kl>} \frac{1}{1+\delta_{kl}} \frac{1}{1+\delta_{ij}}\\
\nonumber
&&  \{ x_1 f_{i/A}(x_1,Q^2) x_2 f_{j/B}(x_2,Q^2)
\left[\frac{d{\hat \sigma}}{d{\hat t}}^{ij->kl}(\hat s, \hat t, \hat u) D_{k->h}(z,\mu^2)
+ \frac{d{\hat \sigma}}{d \hat t}^{ij->kl}(\hat s,\hat u,\hat t) D_{l->h}(z,\mu^2)\right]\\
\nonumber
&&+x_1 f_{j/A}(x_1,Q^2) x_2 f_{i/B}(x_2,Q^2)
\left[\frac{d{\hat \sigma}}{d \hat t}^{ij->kl}(\hat s,\hat u,\hat t) D_{k->h}(z,\mu^2)
\frac{d{\hat \sigma}}{d\hat t}^{ij->kl}(\hat s,\hat t,\hat u) D_{l->h}(z,\mu^2)\right] 
\}
\nonumber
\end{eqnarray}

Details  of the equation can be found in \cite{eskola03}. Here, the parameter
K takes into account the higher order corrections. For the
partonic collisions $ij \rightarrow kl$, $x_1$ and $x_2$ are  the
fractional  momentum  of  the  colliding  partons  $i$  and  $j$,
$x_{1,2}=\frac{p_T}{\sqrt{s}}(e^{\pm  y_1}+e^{\pm  y_2})$,  $y_1$
and $y_2$ being the rapidities of the two final state partons $k$
and  $l$.  For  the parton distribution, $f(x,Q^2)$, we have used
the CTEQ5L parton distribution, with the factorization scale $Q^2
\approx p_T^2$.  
In Eq.\ref{1},  $d{\hat  \sigma}^{ab->cd}/dt$  is
the  sub  process Cross-section. Only 8 sub processes contribute,
they  are  listed  in  \cite{sarcevic}.  $D_h(z,\mu^2)$  is   the
fragmentation function for the final state partons. Fragmentation
functions   are   defined   as   energy   fraction   distribution
$1/\sigma_{total}d\sigma^h/dx$ for the hadron $h$ in the reaction
$e^+ +e^- \rightarrow h+X$. They are not calculable, as formation
of hadrons is non-perturbative. What is done in  practice  is  to
assume a form for the fragmentation function and fit experimental
data  on  hadron production in $e^+e^-$ collisions. Fragmentation
functions are scale dependent. Effect of increasing scale  is  to
shift  the  fragmentation  function  towards  lower  values of z.
Knowing the fragmentation function at some scale, it is  possible
to     obtain     it     at     any     other     scale     using
Gribov-Lipatov-Altarelli-Parisi (GLAP) equation. In  the  present
calculation,  we  have  used KPP parameterization \cite{KPP} with
fragmentation scale $\mu^2 \approx q_T^2$.
The  integration region for $y_2$, $-\ln( \sqrt{s}/p_T -e^{-y_f})
<y_2 <\ln( \sqrt{s}/p_T  -e^{y_f})$,  is  over  the  whole  phase
space, whereas that for z is

\begin{equation}   \frac{2m_T}{\sqrt{s}}   cosh  y  \leq  z  \leq
min[1,\frac{q_T}{p_0} J(m_T,y)] \end{equation}

\noindent with, $J(m_T,y)=(1-\frac{m^2}{m_T^2coshy})^{-1}$. $p_0$
is  the  cut  off  used to regulate infrared singularity. It is a
parameter and in the present work we have used $p_0$=1.0 GeV.

For  AA  collision,  following  the  standard  procedure,  hadron
spectra could be obtained as,

\begin{equation}   \label{2}   \frac{d\sigma^{AA->hX}}{d^2q_Tdy}=
\int_{b1}^{b_2} d^2b d^2s T_A({\bf b-s})T_B({\bf s})
\frac{d\sigma^{NN->hX}}{d^2q_Tdy}
\end{equation}

The  impact  parameter  integration  ranges ($b_1$ and $b_2$) are
chosen according to centrality of collisions.  In  this  picture,
all  the  nuclear  information  is  contained  in  the  thickness
function, $T_A({\bf b})=\int \rho({\bf b},z) dz$.  We  have  used
Woods-Saxon form for the density $\rho(r)$,

\begin{equation}             \rho(r)=\frac{\rho_0}{1+e^{(r-R)/a}}
\end{equation}

\noindent  For  Au,  $R=6.38  fm$  and  $a=0.535 fm$. The central
density $\rho_0$ is  obtained  from  the  normalizing  condition,
$\int \rho(r) d^3r =A$.
 
The key element in  the  jet  quenching or energy loss 
picture  is the fragmentation function $D_h(z,Q^2)$. Unlike in pp
collisions,  in  AA  collisions,  final  state  partons,   before
fragmentation,  can  travel a distance (L) through nuclear matter
and in  the  process  can  lose  energy.  Then  at  the  time  of
fragmentation,  it  has  less energy. For energy loss $\Delta E$,
$z=E_h/E_{parton}$ in the fragmentation function  is  changed  to
$z^*  \rightarrow  \frac{z}{1-\Delta  E/E}$. As the fragmentation
functions are peaked at low z,  with  energy  loss,  fragmentation
will produce less number of hadrons than otherwise. Implementation of
jet quenching in pQCD models is straightforward. While calculating the spectra
for AA collisions, the fragmentation function $D_{i \rightarrow h}(z)$ in Eq.\ref{1}
is replaced such that,

\begin{equation}
z D_{i \rightarrow h}(z)=z^*D_{i \rightarrow h}(z*)
\end{equation}

\noindent with $z^*=z/(1-\Delta E/E)$. 

In a nuclear medium, structure function $f_{a/A}$ is also changed, the so-called EMC effect.
We have taken account of EMC effect through HIJING parameterization. However, its
effect is small.

STAR and PHENIX collaboration measured the nuclear effects on the inclusive spectra, in different centrality ranges via the nuclear modification factor, defined as,

\begin{equation} \label{5}
R_{AA}=\frac{d^2N/dp_Tdy (Au+Au)}{T_{AA}d^2\sigma/dp_Tdy  (pp)}
\end{equation}

\noindent where $T_{AA}=<N_{bin}>/\sigma_{inel}$ from a Glauber calculation
accounts for the nuclear collision geometry. The nuclear modification factor
will be unity, had there been no nuclear effects on high $p_T$ production.
In central collisions, both the
STAR and the PHENIX data show considerable nuclear effects ($R_{AA} < 1$).

For simplicity, we have assumed that in each centrality ranges, partons
suffer on the average $<\Delta E>$ energy.  
Treating the average relative energy
loss $<\Delta E/E>$ as a parameter, we fit the STAR and PHENIX data on nuclear modification factor. The fitting was done using the CERN MINUIT minimizing program. 
 
\section{Results}

In Fig.1a, the $p_T$ spectra for charged hadrons in pp collisions, as obtained by the STAR collaboration is shown.
STAR collaboration counted only 
$\pi^{\pm}$,   $K^{\pm}$  and $p^{\pm}$ as charged hadron and in the calculation
also, we have included those hadrons only. The solid line in Fig.1a, is the fit obtained
to the data in the model. We have fitted only the high $p_T$ part ($p_T>$ 4 GeV)
of the spectra. The LO pQCD give reasonably good description of the data.
In Fig.1b, we have shown the ratio of data to pQCD model calculation. High $p_T$ part
of the spectra is explained within 20\%. We note that better description will
be obtained using generalized parton distribution, however, at the expense of an additional parameter $<k_T^2>$ \cite{ac03}.

In Fig.2, in six panels, we have shown the STAR data on nuclear modification
factor ($R_{AA}$). The solid lines are the fit obtained in the present model. Here also
we have used only the $p_T > 4 GeV$ part of the spectra for the fitting purpose. We note that we are neglecting the Cronin effect. However, at high $p_T$ Cronin effect is small. 
The high $p_T$ part of the STAR nuclear modification factor,
in all the centrality
ranges are well explained.
  pQCD  with jet quenching or the energy loss can
explain the spectacular STAR data on nuclear modification factor,  with a single
parameter, the relative energy loss $<\Delta E/E>$. Later we will discuss in detail
the relative energy loss in STAR data and its implications, for the present, we just mention that
in most central collisions (0-5\% centrality), experimental $R_{AA}$ require that the partons, on the average suffer 20\% energy
loss, ($<\Delta E/E> \sim 0.2$). Energy loss gradually decreases
for less central collisions and   the
most peripheral collisions (60-80\% centrality) are consistent with no energy loss.
Indeed, high $p_T$ part of $R_{AA}$ as obtained by the STAR collaboration in the
centrality range 60-80\% is consistent with no nuclear effect.

As mentioned earlier, we have also analyzed the PHENIX data for the charged hadrons and neutral pions.  In Fig.3 and 4, the nuclear modification factor for the charged hadrons and neutral pions, as obtained by the PHENIX collaboration are shown. The fit obtained in the
model, with the single parameter $<\Delta E/E>$, are also shown in the figures.
As for the STAR data, both the PHENIX data are also well explained in the model. Extracted energy losses also show similar variation with
centrality. 
However, while STAR data require zero or very small energy loss for the most peripheral
collisions, both the  PHENIX data require finite ($\sim$5\%) energy loss even in most 
peripheral collisions. Indeed
as can be seen from the figures, in the most peripheral collisions, while high $p_T$ part of STAR $R_{AA}$ show little or no
nuclear effect, PHENIX data show definite nuclear effect.
Presently we do not understand the reason for this difference. 
 
Extracted average relative energy loss $<\Delta E/E>$ from STAR and PHENIX data are shown in Fig.5,
as a function of number of binary collisions (centrality).
$<\Delta E/E>$ extracted from STAR and PHENIX show similar
centrality dependence. In peripheral collisions energy loss
is less, increasing with the centrality. In most central collisions,
approximately partons on the average suffer $\sim$20\% energy
loss. Extracted $<\Delta E/E>$ from PHENIX neutral pion and
charged hadrons are consistent with each other, but are
consistently larger (5-20\%) than the extracted energy loss from
STAR data. As observed earlier also, even in the most peripheral
collisions, PHENIX data are not consistent with zero or small
nuclear effects, as are the STAR data. In our analysis also, while we obtain nearly zero energy loss for most peripheral STAR collisions, for PHENIX, finite value of energy loss is obtained.

As mentioned earlier, increased energy loss with centrality may have geometric
origin. Simply due to increased path length, partons can suffer more energy loss
in central than in peripheral collisions. 
To understand better the centrality dependence of the extracted energy loss,
using the Glauber model, we have calculated energy lose
suffered by a partons as, 

\begin{equation}
 <\Delta E>({\bf b}) = N({\bf b}) \varepsilon
\end{equation}

\noindent where
$N({\bf b})$ is the average
number of collisions suffered by the partons at impact parameter $b$ and
$\varepsilon$ is the average energy loss in each collisions.
 
At impact parameter  ${\bf  b}$,
the  positions  $({\bf s},z)$ and $({\bf b-s},z\prime)$ specifies the initial hard scattering of partons. The number of  collisions
suffered  by  a  parton  after the initial hard scattering can be
calculated as,

\begin{equation}
N({\bf b,s},z,z\prime) = \sigma  \int_z^\infty dz_A \rho({\bf s},z_A) +\sigma 
\int_{z\prime}^\infty
dz_B  \rho({\bf b-s},z_B)
\end{equation}

\noindent where $\sigma$ is the parton-nucleon cross section.
Above expression should be averaged over  all  positions
of   initial  hard  scattering  with  a  weight  of  the  nuclear
densities.

\begin{equation}
N(b)=\int d^2s \int dz \rho_A(s,z) \int dz\prime \rho_B(b-s,z\prime) N(b,s,z,z\prime)/
\int d^2s \int dz \rho_A(s,z) \int dz\prime \rho_B(b-s,z\prime)
\end{equation}

In Fig.5, the solid line is the Glauber model  calculation of energy loss (suitably
normalized). 
We note that simply due to geometry of the collisions, even in cold nuclear matter, energy loss depend on the centrality.
In central collisions, the partons can travel a distance longer than in peripheral
collisions, leading to more energy loss in central than in peripheral collisions.
It will be unrealistic to associate this increased energy loss in central collisions
with the medium density. Naturally, the energy loss extracted from the STAR and PHENIX data include the geometric effect and only by eliminating the geometric
effect, can one comment on the nature of the medium. 
We note that Glauber model calculation cannot reproduce the centrality
dependence of the extracted energy loss in all the centrality ranges. While in
peripheral collisions, Glauber model can account for the extracted
energy loss, in more central collisions, energy loss is underpredicted by a factor of 1.5-2.0.
In peripheral collisions, we do not expect any other medium than normal nuclear
matter. Agreement of Glauber model calculation with the extracted energy loss 
in peripheral collisions then
suggests that in a
peripheral collision, possibly normal nuclear matter is produced. 
On the same reasoning,  failure
of the Glauber model to account for the energy loss in central collisions
indicate matter other than normal nuclear matter is produced in central collisions. 
 For  central collisions, Glauber model predict 1.5-2.0 times less energy
loss than the extracted values. Energy loss in excess of Glauber model calculation
can be attributed to dynamical origin. If we assume that energy loss is
proportional to the  density, this would lead to 1.5-2.0 time higher 
density in central collisions than in peripheral collisions. This is much less
than the density (15 times normal density) obtained by Wang and Wang \cite{wa02}.
The present analysis thus precludes any high density formation in RHIC Au+Au collisions.

Partons lose energy mainly due to gluon radiation. In a static medium, energy
loss due to gluon radiation exhibit a characteristic quadratic dependence on the
in medium path length L, $\Delta E \propto L^2$. For a one dimensionally expanding medium, the
quadratic dependence gets diluted to a linear dependence \cite{gy00a}.
In Fig.6, we have plotted the extracted energy loss as a function of the in medium
path length (L). Fit to the
extracted energy loss with the parametric form, 
$<\Delta E/E> = A L^2$ are also shown in Fig.6. 
Energy loss from the STAR data are well explained with
$A=6.24\times10^{-3} fm^{-2}$ (the solid line). For the PHENIX experiment also,
$<\Delta E/E>$ in more central collisions are explained by the
quadratic form with $A=7.41\times10^{3} fm^{-2}$, but under predict
the energy loss in peripheral collisions. 
 Quadratic dependence of the energy loss in central collisions
on the in medium path length then suggests that partonic energy
loss is mainly due to gluon radiation in a {\em static medium}, i.e. they lose
the energy before the expansion sets in.
Hydrodynamic expansion sets   in a time scale of 0.2-1.0 fm. The hadrons then
are formed before that time scale. It is very important result. If hadrons are
formed in a time scale of 0-2-1.0 fm, they can possibly be absorbed in the
resulting hadronic medium. They high $p_T$ suppression then may be due to the
combined effect of jet quenching and hadronic interaction in medium.
 
\section{Summary and Conclusions}

To summarize, we have analyzed the STAR and PHENIX data on high $p_T$
suppression in Au+Au collisions at   $\sqrt{s}$=200 GeV.
Assuming that in each centrality range, partons suffer on the average $<\Delta E>$ energy loss, we have extracted the relative energy loss $<\Delta E/E>$
from a fit to the experimental nuclear modification factor. Energy loss thus extracted show that partons suffer more
energy loss than in central than in peripheral collisions. However, only a part
of the energy loss is of dynamical origin. A major part of the centrality
dependence of energy loss is geometric, i.e. due to increased path length in central
than in peripheral collisions. Extracted energy loss in peripheral
STAR and PHENIX collisions are consistent with the energy loss calculated
in a Glauber model. However, Glauber model predict 1.5-2.0 times
less energy loss in central collisions than the extracted values. The excess energy
loss is of dynamical origin. If the energy loss is proportional to the density, then
present analysis indicate that in central collisions, density of the medium is
1.5-2.0 times higher than that of normal nuclear matter. Our analysis thus precludes
any high density matter formation. We have also investigated the in medium
path length dependence of the extracted energy loss. Energy loss in STAR and
also PHENIX in central collisions show a quadratic dependence on the
medium path length, characteristic of gluon radiation in a {\em static} medium.
It appears that partons lose their energy  before the hydrodynamic
expansion sets in. 

To Conclude, our analysis of STAR and PHENIX data on high $p_T$ suppression indicate 
that in central collisions, partons lose energy due to gluon radiation, in a static medium, of density, 1.5-2.0 times normal nuclear density.

\eject
\begin{figure}[h]
\centerline{\psfig{figure=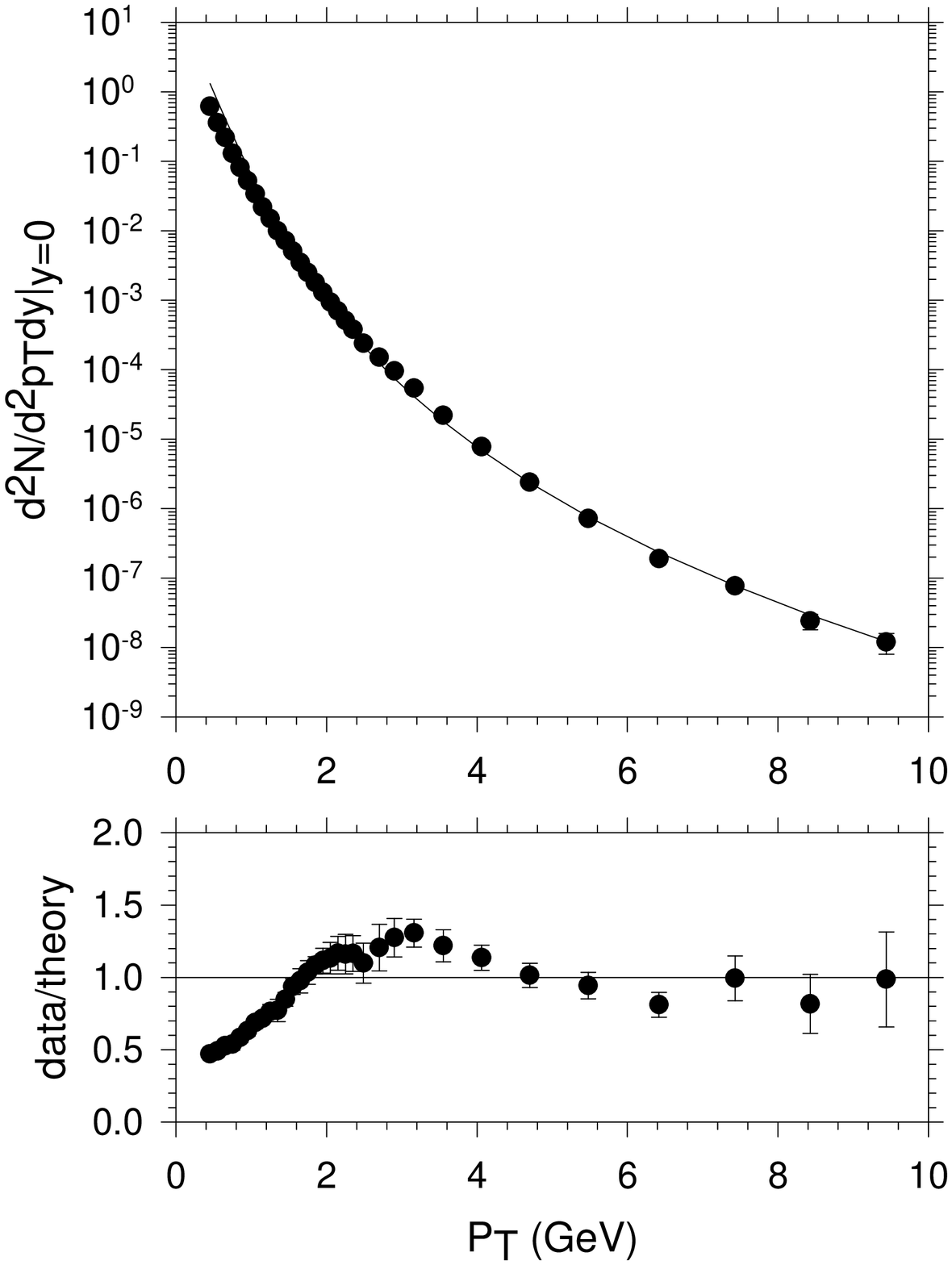,height=10cm,width=7.5cm}}
\caption{(a)The  transverse  momentum  spectra  for the
charged hadrons, in STAR pp collisions. The solid line is the pQCD calculation. 
(b) The ratio of the data to the theory.}
\end{figure}

\begin{figure}[h]
\centerline{\psfig{figure=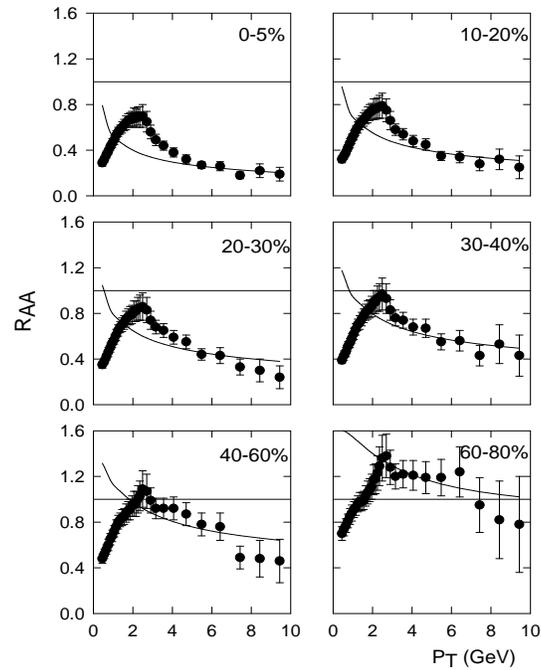,height=10cm,width=7.5cm}}
 \caption{The STAR data on nuclear modification factor for the charged hadrons and the fit obtained to the data in the present model.}
\end{figure}

\begin{figure}[h]
\centerline{\psfig{figure=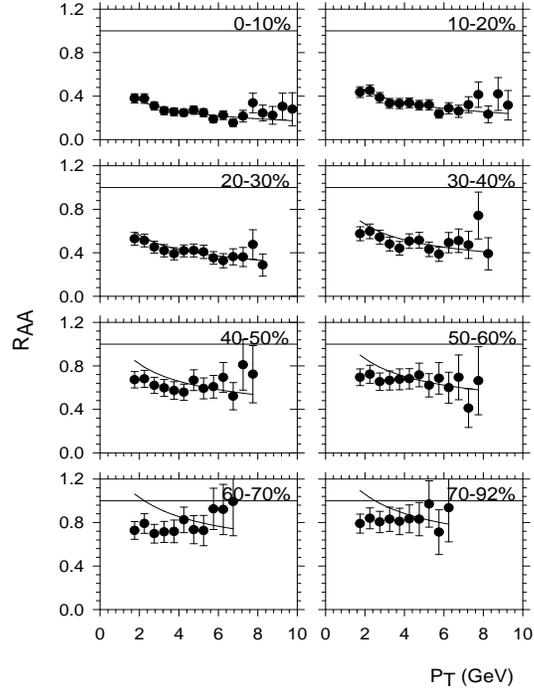,height=10cm,width=7.5cm}}
 \caption{Same as Fig.2 for PHENIX $\pi^0$.} 
\end{figure}

\begin{figure}[h]
\centerline{\psfig{figure=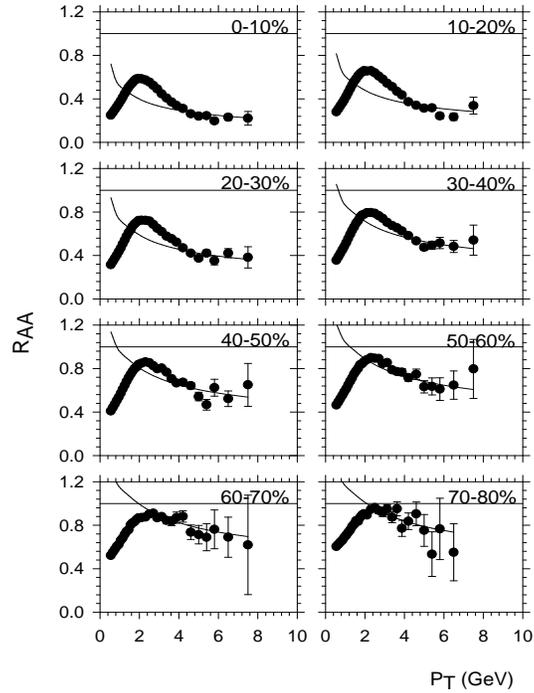,height=10cm,width=7.5cm}}
 \caption{Same as Fig.2 for PHENIX charged hadrons.}
\end{figure}

\begin{figure}[h]
\centerline{\psfig{figure=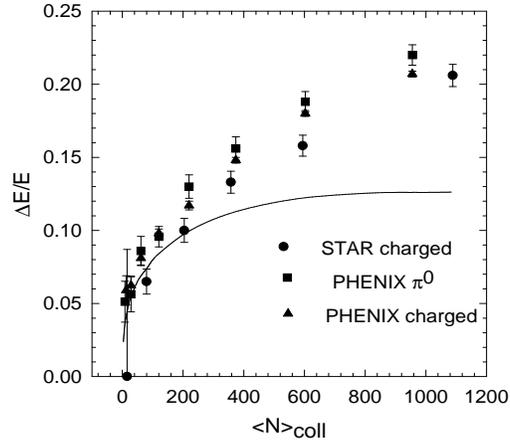,height=10cm,width=7.5cm}}
\vspace{-2cm} 
 \caption{ The extracted relative energy loss
as a function of number of binary collisions (centrality of collisions). The solid line is a Glauber model calculation of
energy loss in nuclear matter.}
\end{figure}

\begin{figure}[h]
\centerline{\psfig{figure=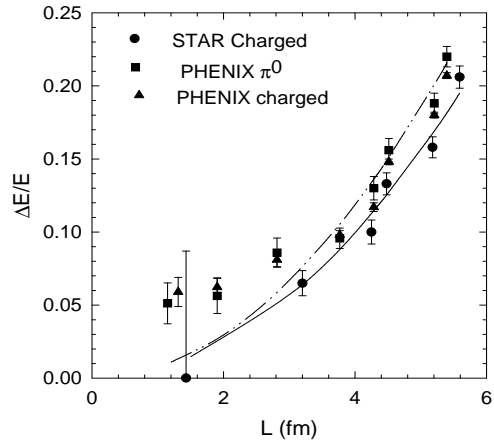,height=10cm,width=7.5cm}}
\vspace{-2cm} 
 \caption{ The extracted relative energy loss as
a function of in medium path length. The solid and dashed lines are fit to
the extracted values from the STAR and PHENIX data, with a quadratic L dependence.}
\end{figure}
 
\end{document}